\documentclass[a4paper,fleqn,usenatbib]{mnras}
\usepackage{epsfig,epstopdf,color,ulem,appendix}

\newcommand{\dotm}{\dot{m}}
\newcommand{\rg}{{\rm R_g}}

\title[SED of the inner accretion flow around Sgr A*]{Spectral Energy Distribution of the inner accretion flow around Sgr A* -- Clue for a weak outflow in the innermost region}

\begin{document}
\author[Ma et al.]{
Ren-Yi Ma$^{1,2,5}$ \thanks{ryma@xmu.edu.cn},
Shawn R. Roberts$^{2,3}$ \thanks{roberts.shawn513@gmail.com}, 
Ya-Ping Li$^{4,5}$ \thanks{liyp@shao.ac.cn}, 
Q. Daniel Wang$^{2}$\thanks{wqd@umass.edu}\\
$^{1}$ Department of Astronomy and Institute of Theoretical Physics and Astrophysics, Xiamen University, Xiamen, Fujian 361005, China \\
$^{2}$ Department of Astronomy, University of Massachusetts, Amherst, MA 01002, USA\\
$^{3}$ Emprata LLC, Clifton, VA 20124, USA \\
$^{4}$ Key Laboratory for Research in Galaxies and Cosmology, Shanghai Astronomical Observatory, Chinese Academy of Sciences, 80 \\
$~~$Nandan Road, Shanghai 200030, China\\
$^{5}$ SHAO-XMU Joint Center for Astrophysics, Xiamen, Fujian 361005, China
\vspace{-0.5cm}}

\maketitle

\label{firstpage}
\begin{abstract}
Sgr A* represents a unique laboratory for the detailed study of accretion processes around a low-luminosity supermassive black hole (SMBH).  
Recent X-ray observations have allowed for spatially resolved modeling of the emission from the accretion flow around the SMBH, placing tight constraints on the flux and spectral shape of the accretion from the inner region with $r<10^3~\rg$, where $\rg \equiv GM_{\rm BH}/c^2$ is the gravitational radius of the black hole with mass of $M_{\rm BH}$.
We present here the first modeling of the multi-band spectral energy distribution (SED) of this inner region to better constrain the physical condition of the innermost accretion flow.
Our modeling uses the Markov chain Monte Carlo (MCMC) method to fit the SED,
accounting for the limitations on the accretion rate at the outer radius of $10^3~\rg$ from the earlier works and the domination of the accretion flow within $30~\rg$ to the sub-mm bump.
It is found that the fitting results of the outflow index could be very different.
If only the most luminous part of the SED, the sub-mm bump, is considered, the outflow index is about 0, while if low-frequency radio data and X-ray data are also included, the outflow index could be 0.37 or even higher.
The great difference of the fitting results indicates that the outflow index should be variable along radius, with a strong outflow in the outer region and a weak outflow in the innermost region.
Such weak outflow agrees with numerical simulations and is possible to explain the multi-band SED even better.
\end{abstract}

\begin{keywords}
accretion, accretion discs - Galaxy: centre
\vspace{-0.5cm}
\end{keywords}

\section{Introduction}

The proximity of the Galactic Center has made Sgr A* an important and unique laboratory for us to understand the accretion process around black holes (BHs), even though it is quite dim with most of the luminosity being radiated in the sub-mm bump. 
Spurred by considerable observational data, the general picture of inflow and outflow around Sgr A* has begun to emerge.  
As stellar winds spew forth from Sgr A*'s massive, orbiting stars, they collide with each other, shocking to X-ray emitting temperatures (e.g., \citealt{Quataert05,Cuadra2008,Cuadra2015}).  Part of this gas is captured by the SMBH at the center of Sgr A* and begins falling deeper into the potential well.  With a substantial amount of coherent angular momentum, the gas is at least partially rotationally supported throughout \citep{Roberts2017}.  
As angular momentum is transported, the gas turbulently funnels inward.  
Some of this gas will accrete onto the BH, while most of it is driven away in a strong, collimated, polar outflow \citep{Yuan12,Narayan12,Li2013,Wang2013,Yuan2015,Roberts2017}.  
In general, we expect this understanding to be readily extendible to other low luminosity active galactic nuclei (LL-AGN), weakly accreting and emitting supermassive black holes (SMBHs).
Thus, a detailed study can help us to broadly understand how feedback from these objects imposes on the circumnuclear environment \citep{Gan14,Li18}.  
However, even with this general framework in place, some details of the accretion flow remain to be understood,
one of which is the outflow in the region close to the BH.

The outflows/winds from hot accretion flows have been found in numerical simulations for some time \citep[e.g.,][]{Stone99,I99,Ohsuga11}.
\citet{Stone99} found that both the inflow rate and the outflow rate decrease inward, roughly following a power-law behavior, with the accretion rate $\dotm(r) \propto r^{s}$ and the density $n(r) \propto r^{-3/2+s}$.
The typical value of $s$, the outflow index, is $0.5-1$ from the simulations \citep{Yuan12}. 
The outflow origin has been studied via several different ways.
\citet{Begelman12} analytically studied the outflow from the conservation laws of 
energy and angular momentum.
The existence of outflow was also proposed in \citet{Gu15} based on
stability analysis. 
By tracking the trajectories of virtual 
particles, \citet{Yuan2015} not only show the existence of the outflow, but 
also calculated the detailed properties of the outflow such as their mass flux 
and velocity based on general relativistic magneto-hydro dynamic (GRMHD) numerical simulations. They have also analyzed 
the driving mechanism of the outflow, which is the combination of magnetic 
and centrifugal forces.

\begin{figure}
\centerline{\epsfig{figure=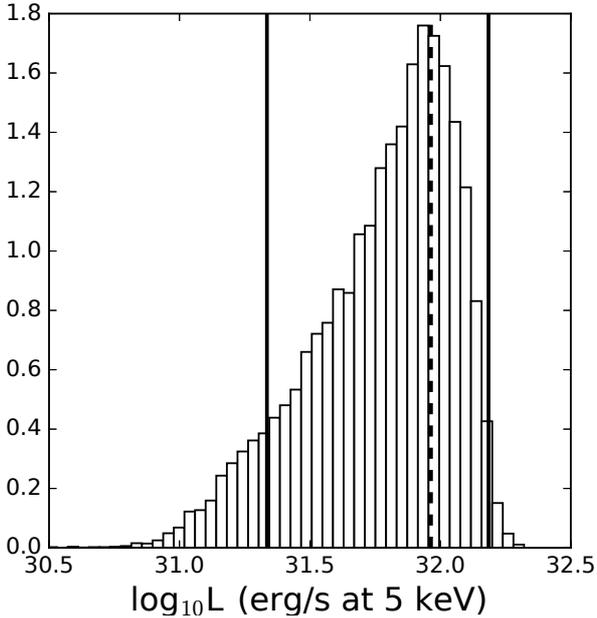,width=0.5\textwidth,angle=0}}
\caption{
The probability distribution of the point-like flux (egs/s) at 5 keV.}
\label{f:ps_dist}
\end{figure}

\begin{figure}
\centerline{\epsfig{figure=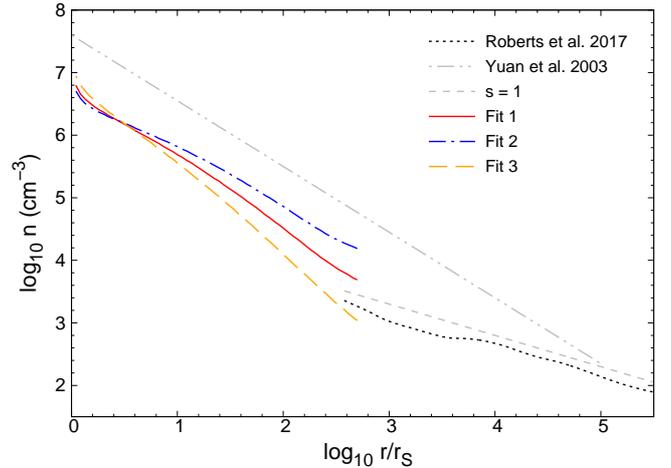,width=0.5\textwidth,angle=0}}
\caption{
The electron density as a function of radius.  
The black dotted line shows the best fit azimuthally averaged profile found by Roberts et al. (2016).  The Roberts profile appears to be much shallower than that of Yuan, although this comparison is somewhat problematic, because the outer boundaries of the flow inferred or assumed are quite different in these two cases.
Furthermore, the modeling in Y03 is dominated by the SED in the radio-NIR range, which arises exclusively in the inner region of the accretion flow, while the analysis in \citet{Roberts2017} is based chiefly on the extended X-ray emission from the outer region. 
}
\label{f:density}
\end{figure}

Significant outflows from the hot accretion flow around the central BH in Sgr A* have been supported by observations.
First, the accretion rate at Bondi radius is found to be $\sim 10^{-5} M_{\odot}~{\rm yr^{-1}}$, however the bolometric luminosity is quite low $\sim 10^{36} {\rm ergs~s^{-1}}$ \citep[e.g.,][]{Genzel94,Baganoff03}. 
The radiatively inefficient accretion flow (RIAF) model without outflow can account for the low luminosity but overestimates the rotation measure (RM) by many orders of magnitude \citep{Quataert2000}. 
In contrast, the RIAF model with outflow can explain both the luminosity and RM \citep[][]{Yuan2003}.
Second, recent observations by \citet{Wang2013} and \citet{Roberts2017} have found the H-like Fe~K$\alpha$ line is weak and the X-ray flux is spatially flat, which requires flat radial density distribution and indicates the outflow index as high as what numerical simulations showed, i.e., $s\sim 1$.

Numerical simulations have also shown that there is a characteristic radius $\sim 30-90~\rg$,
beyond which the outflow dominates while inside the inflow dominates \citep[e.g.][]{Stone99, Yuan12b,Narayan12, McKinney12,Li2013,Yuan2014}.
The domination of inflow in the innermost region is easy to understand, because the gravitational potential close to the BH is so steep that the accretion timescale beomes shorter than the timescale required for the formation of outflow \citep{Yuan12}.
However, at present, the region where the observations have constrained the outflow rate well is still quite large, $\ge 10^3~\rg$.
It is interesting to explore the region of smaller radius, or even the innermost region of the accretion flow.

Detailed multi-band SED modelling or fitting could provide us with information about the accretion flow in the inner region.
One important study for understanding the quiescent emission of the Sgr A* accretion flow was that by \citet[][, hereafter Y03]{Yuan2003}.  
The authors explained the multi-wavelength SED with a RIAF model, in which the outflow index is $s\sim 0.3$.
The sub-mm bump was explained by the synchrotron radiation of the thermal electrons, 
the X-ray flux was mainly from bremsstrahlung radiation in the outer region of the RIAF,
and the radio emission at low frequencies was explained by non-thermal electrons.
This model successfully explained the SED, as well as the observed RM.
It was even consistent with newer estimates of the mean IR flux \citep{Schodel2011}. 
However, the region they explored was large ($r \leq 10^5~\rg$).
Since the X-ray emission was dominated by the outer parts of RIAF, the X-ray data they used could not impose strong constraints on the inner accretion flow.

Since the publication of Y03, significant progresses in radio and X-ray observations have been made, enabling us to understand the accretion flow better.
New radio observations are availiable \citep[][and references therein]{Shcherbakov12}, especially those observations by ALMA \citep{Liu16a, Liu16} in mm/sub-mm band that 
dominates the luminosity of Sgr A*.
The 3-megaseconds {\it Chandra} X-ray observation of quiescent Sgr A* provides an unprecedent diagnostic capability to probe the accretion flow \citep{Wang2013}.
By comparing three different band images of the combined quiescent {\it Chandra} data to numerical simulations,
\citet{Roberts2017} deconvolved the residual point-like emission of the gas very close to the BH ($r<10^3~\rg$) via the power of MCMC sampling.  
They found that the point-like component had a specific luminosity of log$_{10}\nu$L$_\nu\sim31.96$ (31.32, 32.18) ergs/s at 5 keV, and was responsible for 4.2\% (2.3, 7.0) of the observed emission within 1.5" in the 1-9 keV band (see Figure~\ref{f:ps_dist}).  
The authors also found that the point-like emission could be well characterized by a single power-law, with $\alpha\sim4.8$ (3.5, 7.5; 90\% confidence interval). 
Comparing with the X-ray data used in Y03, the contribution of the inner accretion flow to the X-ray flux could be much better constrained.

\begin{figure*}
   \centering
  \includegraphics[width=0.8\textwidth, angle=0]{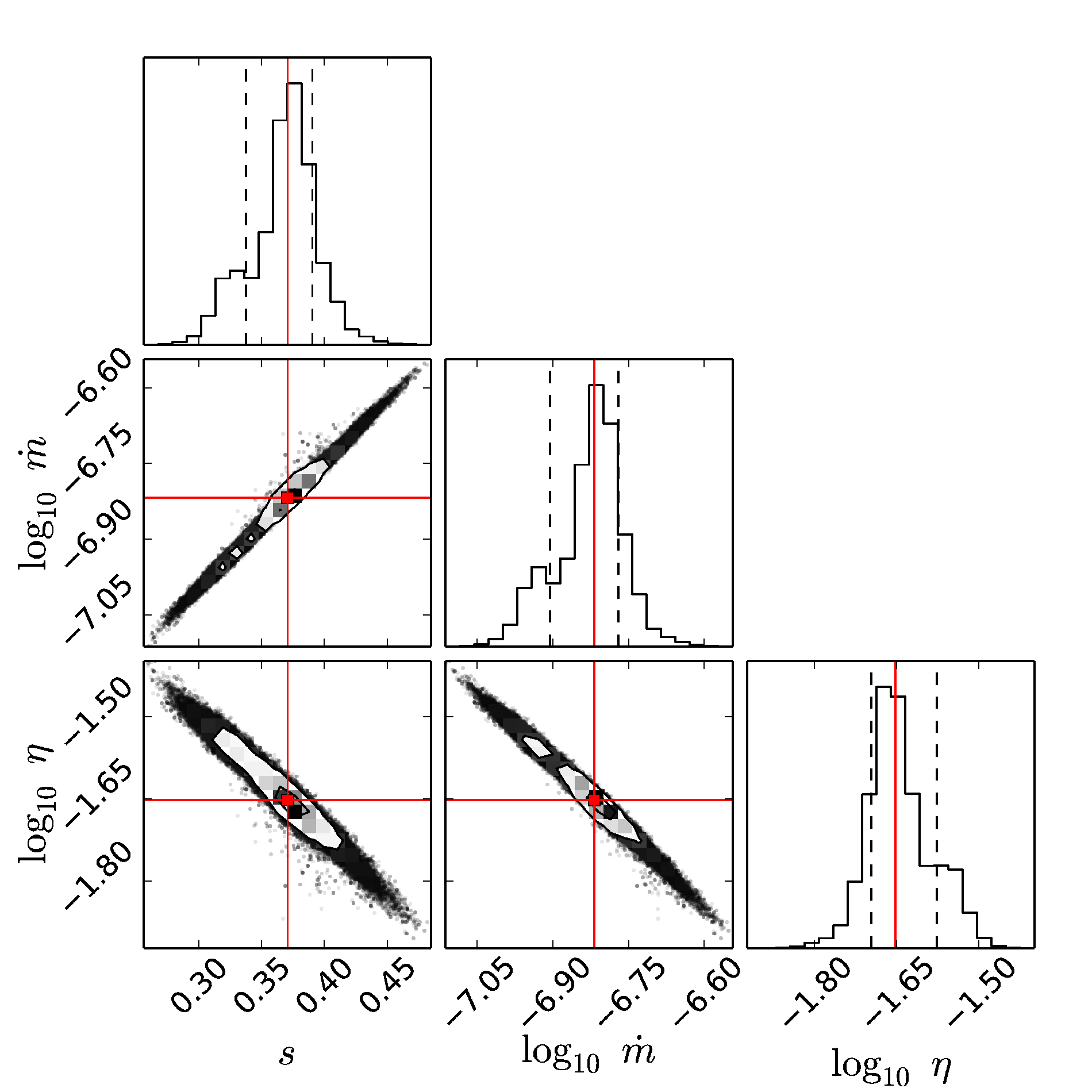}
\caption{Contours of MCMC samplings in Fit 1. The countour lines show the region of $2\sigma$ and $1\sigma$ levels. The red lines in the 1D histograms show the mean value and the dashed lines show the range of $1\sigma$ deviation.}
\label{fig:triangle1}
\end{figure*}

\begin{figure}
   \centering
  \includegraphics[width=0.4\textwidth, angle=0]{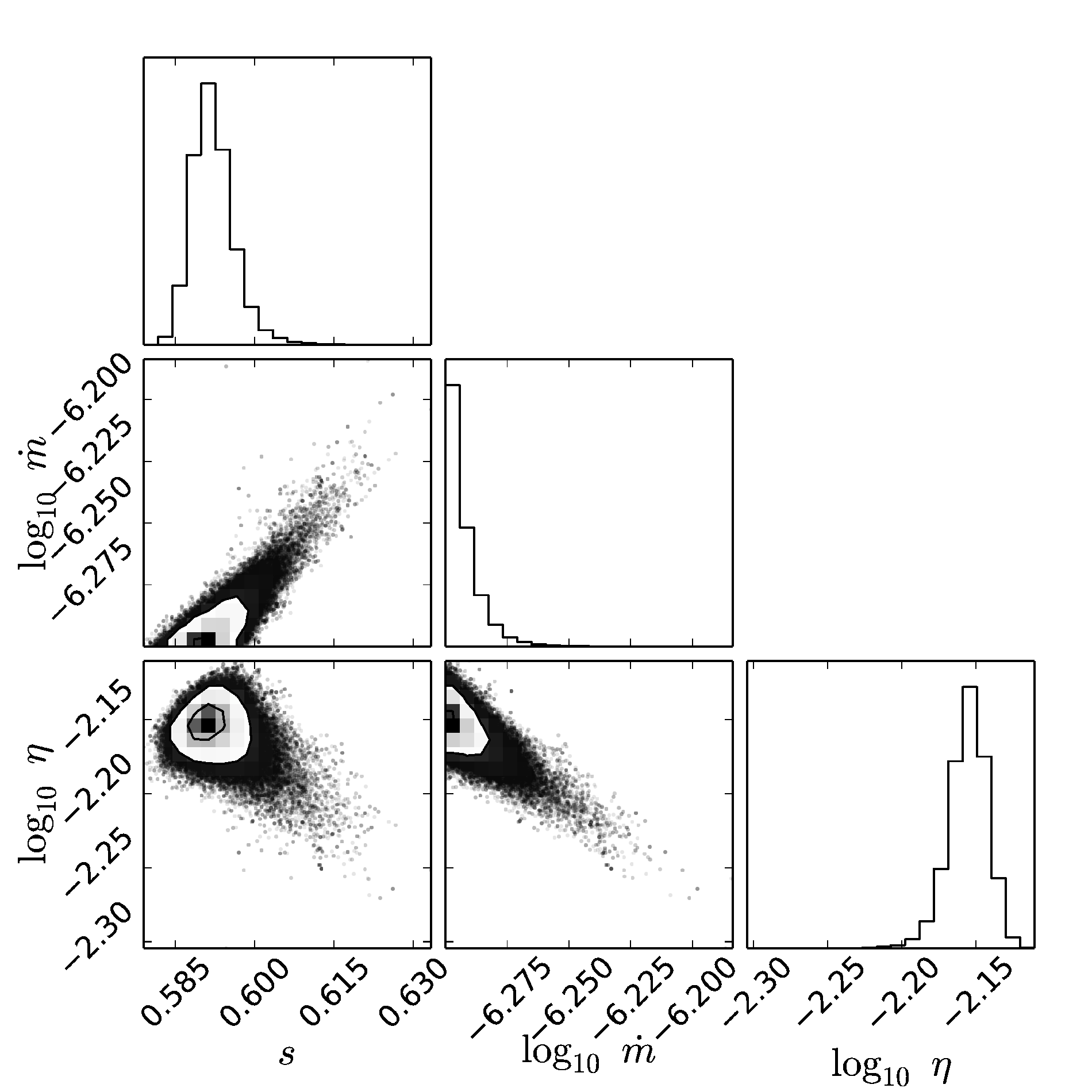}
\caption{Contours of MCMC samplings in Fit 2. The countour lines show the region of $2\sigma$ and $1\sigma$ levels. }
\label{fig:triangle2}
\end{figure}

\begin{figure}
   \centering
  \includegraphics[width=0.4\textwidth, angle=0]{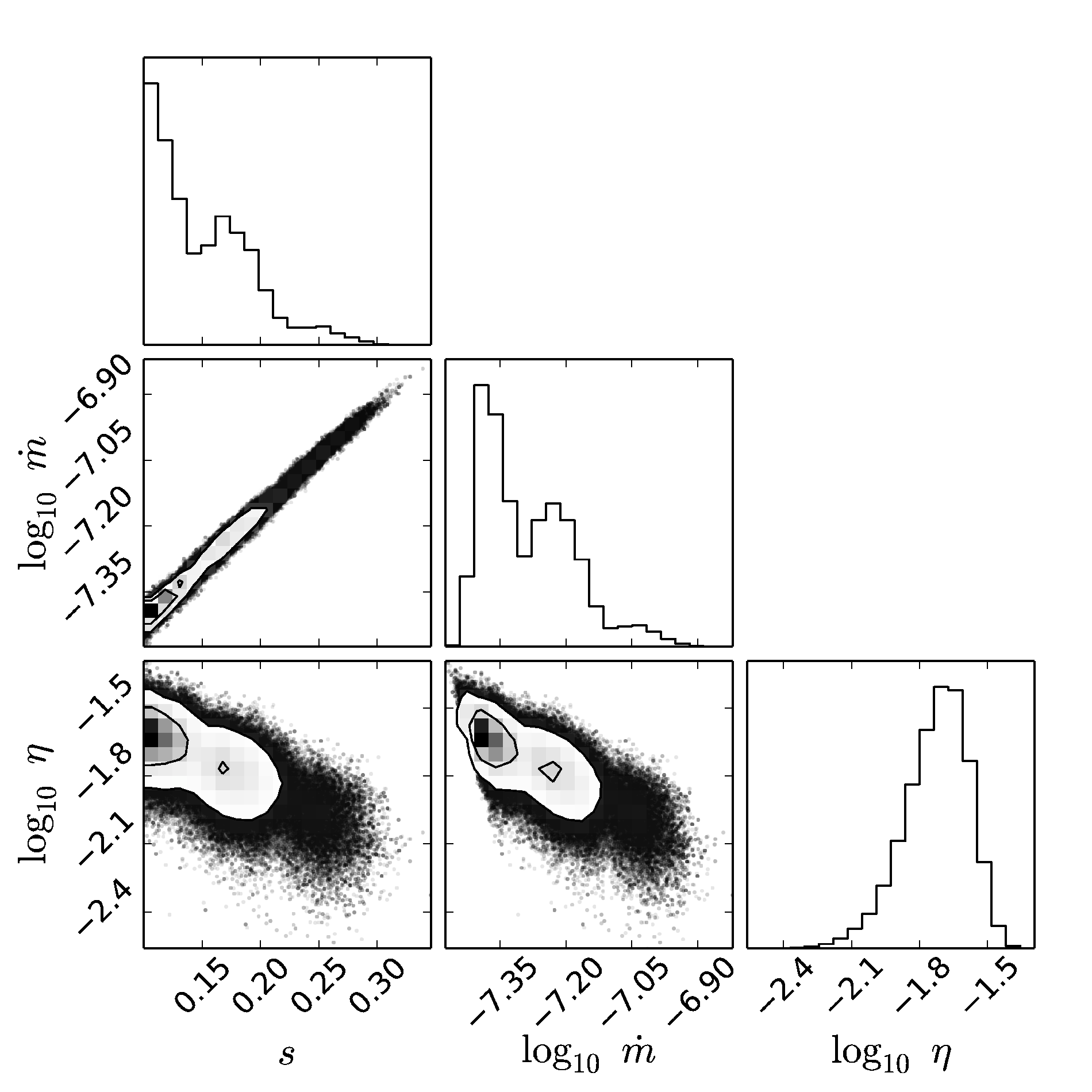}
\caption{Contours of MCMC samplings in Fit 3. The countour lines show the region of $2\sigma$ and $1\sigma$ levels.}
\label{fig:triangle3}
\end{figure}

Considering these recent advances, in this paper, we re-investigate the multi-wavelength SED.
We only consider the inner region within $10^3~\rg$  and leave the index $s$ a free parameter, which could be different to the value in the outer region of $r>10^3~\rg$ ($s\sim 1$).
Moreover, instead of just modeling the data via intuitive parameter adjustments, MCMC is  used for the first time to fit the multi-band SED. 
The fit allows us to infer the state of innermost accretion flow and to estimate the uncertainties in the key model parameters.

\section{Methods}
\label{sec:methods}

We calculate the emergent SED following the procedure described in Y03.
First we calculate the one-dimensional dynamics of the RIAF by solving for the conservation equations of mass, radial momentum, angular momentum, and energy equations for electrons and ions, i.e., Equations 1-5 in Y03,
\footnote{It should be noted that these equations are not fully consistent since they do not account for energy or angular momentum that is lost in an outflow. 
However, an outflow primarily manifests itself through a flattening of the density profile, which is allowed for (e.g., \citealt{Xie2008}).} 
with the boundary conditions of the electron/ion temperature and mass accretion rate, which are constrained by recent results \citep{Wang2013,Roberts2017}. 
And then  we calculate the output spectrum, taking into account the synchrotron emission of the thermal and non-thermal electrons, their inverse Compton scattering, and the bremsstrahlung of the thermal electrons.
It should be noted that following \citet{Yuan06}, the density and temperature of the inner accretion flow is modified according to the result of \citet{Popham98} as the Paczynski-Witta pseudo-potential is used in Y03.
This is not very precise, but it can take into account the general relativistic potential acceptably.

It should be mentioned that GRMHD numerical simulations show coupled disk-jet structure, and the detailed radiative transfer calculations successfully explained the sub-mm bump and flat radio spectra \citep[e.g.,][]{Dexter10,Moscibrodzka13,Moscibrodzka14,Ressler17}. 
However, whether the jet exists in Sgr A* or not is still up for debate.
Detailed calculations have shown that the SED observed should not arise from the jet, because the predicted RM was at least two orders of magnitude smaller than that observed \citep{Li15}.
So even if a jet exists, it should be very weak. 
Therefore, the jet is ignored in this paper.

The BH mass has been well determined from the stellar orbits as $4.1\times 10^6 M_\odot$ \citep[e.g.,][]{Genzel10}, and the distance is $\sim 8$~kpc.
The radius of concern ranges from $\sim 2~\rg$ to $10^3~\rg$.
There are other parameters in the RIAF model.
We fix some parameters at their typical values, with the viscous parameter $\alpha=0.1$, the ratio of the gas pressure $p_{\rm gas}$ to the magnetic pressure $p_{\rm mag}$, i.e., $\beta\equiv~p_{\rm gas}/p_{\rm mag}=10$ \citep{Yuan2014},
the fraction of the turbulent energy heating the electrons $\delta=0.3$,
and the energy spectral index of the non-thermal electrons $p=3.5$.
Three key parameters are left for fitting the SED, i.e.,
the accretion rate at the outer boundary $\dotm_0=\dotm(r=10^3~\rg)$, the fraction of the non-thermal electrons to the thermal electrons $\eta$, 
and the index of the mass outflow $s$.

The multi-band SED data are collected from the published papers: radio \citep{Shcherbakov12,Liu16a,Liu16}, IR \citep{Schodel2011}, and X-ray \citep{Roberts2017}, 
all of which are mainly from the inner region ($r<10^3~\rg$).
It should be noted that in each of the two narrow frequency ranges around $\sim 700 {\rm GHz}$ and $\sim 10^{14} {\rm Hz}$, 
there are more than one data points with the fluxes being different almost as much as twice.
The varaiation of the flux should not arise from the intrinsic radiation of the quiescent state, as the timescale should be much longer.
Probably they are related with the flares, the radiations of which superimpose on the intrinsic quiescent emission.
So in our fittings, we only select the lowest flux in these two narrow frequency ranges.

Instead of simply modeling,
we use the MCMC method to fit the SED, which gives the distribution of probabilities for the three fitting parameters.
We use the {\it emcee} code to do MCMC sampling, which is available online at {\it http://dan.iel.fm/emcee}. 
The script '{\it Corner.py}' is used to plot the contours of MCMC samplings \citep{corner}. 
The logarithmic likelihood function is 
\[
{\rm ln}~p(L|\nu,s,\dotm_0,\eta)=-\frac{1}{2}\sum\limits_{i=1}^N\frac{(L_i-L_{{\rm model},i})^2}{\sigma_i^2},
\]
where $N$ is the total number of observational data, $L_i$ is the observed flux at frequency $\nu_i$, $L_{{\rm model},i}$ is the luminosity calculated from the RIAF model, and $\sigma_i$ is the error bar of each data.

Since computing the RIAF spectra is time consuming, interpolation is used to obtain the spectra.
First, two tables of spectra for different grids of $s$ and $\dot{m}_0$ are calculated.
One is for thermal electrons, while the other is for non-thermal electrons.
Because the contribution of the non-thermal electrons at any frequency is proportional to $\eta$, 
the table of spectra is calculated with $\eta=0.01$. 
While for any other value of $\eta$, the spectra of non-thermal electrons can be easily calculated by multiplying $100\eta$.
Then combine the spectra of thermal and non-thermal electrons, 
the spectra at different grids of $s$ and $\dot{m}_0$ for any $\eta$ is obtained.
And finally, the spectra for any value of $s$ and $\dot{m}_0$ is derived by interpolation.
In our calculations, the value of $s$ ranges from  0.0 to 1.0 with grid of 0.02, and $\rm{log_{10}}\dot{m}_0$ ranges from -7.5 to -5.5 with grid of 0.2.

\begin{table*}
\centering
\caption{Fitting parameters to the multi-band SED. For Fit 2 and Fit 3, the values of the parameters are taken at the peaks of the histogram as shown in Figures~\ref{fig:triangle1} \& \ref{fig:triangle3}. The modeling results of Y03 are also shown for comparison.}
\label{tab:para}
\resizebox{0.8\textwidth}{!}{%
\begin{tabular}{ccccc}
\hline\hline
  & Fit 1 & Fit 2 & Fit 3 & Y03 \\ \hline
&&&&\\
differences & N/A & limited range of $\dotm_0$ & only submm \& IR & $r<10^5~\rg$ \\
&&&&\\
$\dotm_0~(10^{-7} M_{\odot}~{\rm yr^{-1}})$ & 1.5 & 5.0 & 0.3 & 2.4  \\
&&&&\\
$s$ & 0.37 & 0.59 & 0.05 & 0.27 \\
&&&& \\
$\eta$ (\%) & 2.2 & 0.7 & 1.9 & 1.5 \\
\hline\hline
\end{tabular}
}
\end{table*}

\begin{figure}
\centerline{
\epsfig{figure=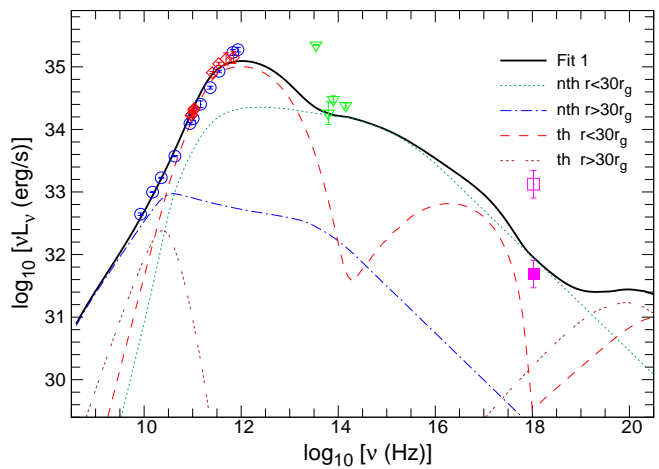,width=0.5\textwidth,angle=0}}
\caption{ 
The SED of Fit 1 to the point-like source in Sgr A*. The contribution of thermal and non-thermal electrons in the inner ($r < 30~\rg$) and outer ($30~\rg < r < 10^3~\rg$) regions to the SED are also shown with different lines.
For the thermal electrons in the inner region, the three bumps from low to high frequencies correspond to synchrotron, inverse Compton scattering, and bremsstrahlung emission, respectively.
For the thermal electrons in the outer region, the bump of inverse Compton scattering is too weak to be shown.
The origins of the data in both panels are shown with different symbols of circles~\citep{Shcherbakov12}, diamonds~\citep{Liu16}, downward triangles~\citep{Schodel2011}, empty square~\citep{Wang2013}, and filled square~\citep{Roberts2017}.
}
\label{fig:sedfit}
\end{figure}

\begin{figure*}
\centerline{
\epsfig{figure=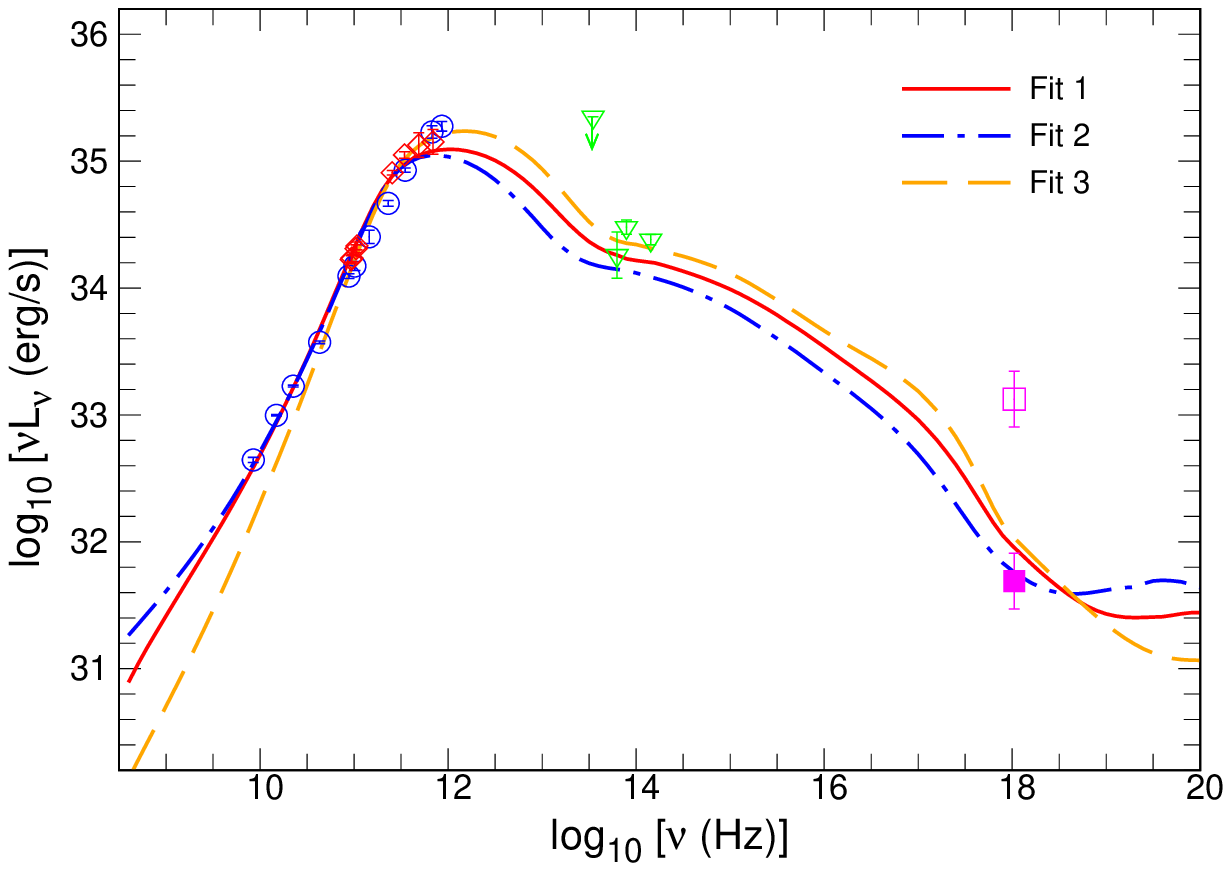,width=0.5\textwidth,angle=0}
\epsfig{figure=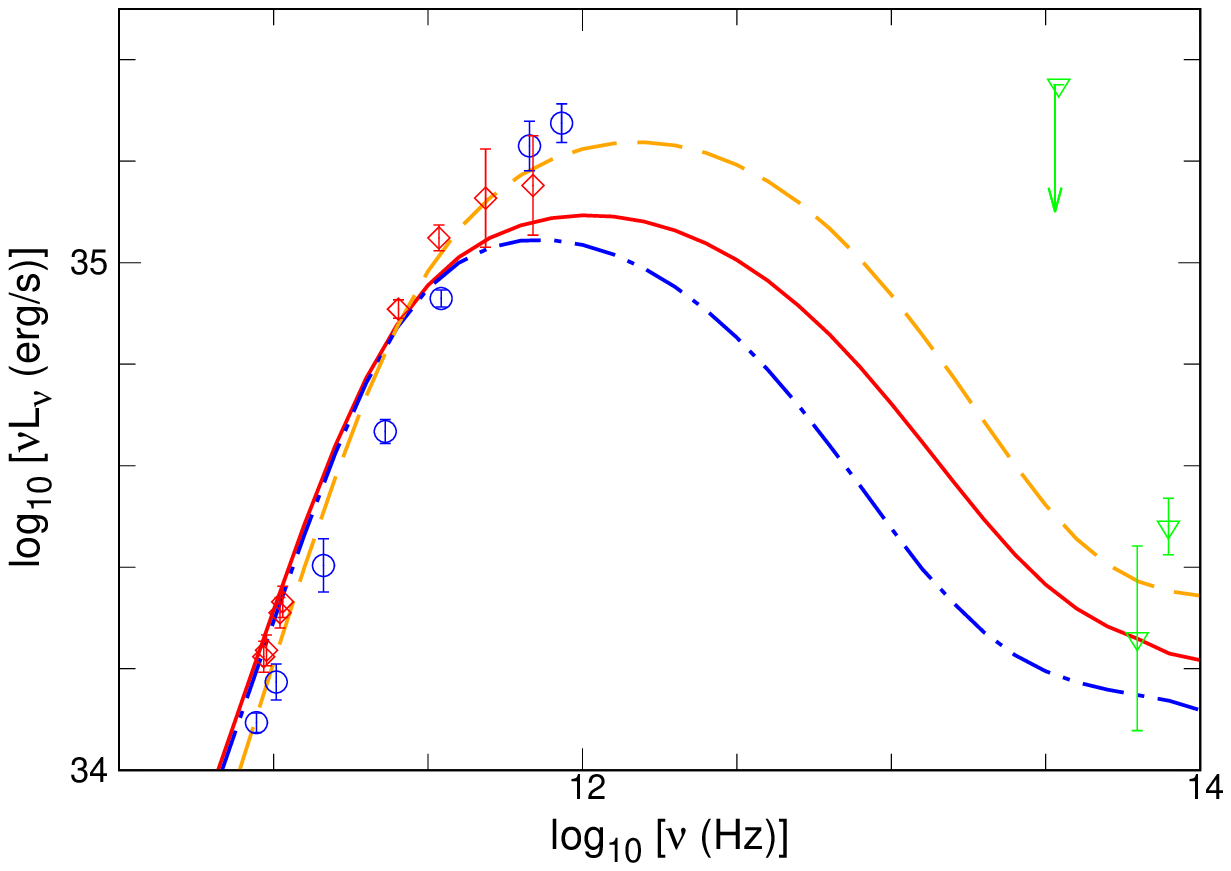,width=0.5\textwidth,angle=0}}
\caption{ 
{\it Left panel ---}The SED of different fittings. 
The solid, dot-dashed, and dashed lines correspond to the fittings with unconstrained accretion rate, constrained accretion rate, and only sub-mm bump data, respectively.
{\it Right panel ---}Zoom in of the left panel around the sub-mm bump.
The origins and captions of the data in both panels are the same as Figure~\ref{fig:sedfit}.
}
\label{fig:compa}
\end{figure*}

With the tabulated spectra, we made three different fittings in this paper, i.e. Fits 1-3.
In Fit 1, we fit the multi-band SED without any limitation to the parameters.
Considering the limitation on the accretion rate at the outer boundary from \citet{Roberts2017}, $\sim 10^{-6} M_{\odot}~{\rm yr^{-1}}$, we made Fit 2 within a very limited range of the accretion rate, i.e. $0.5-2\times 10^{-6} M_{\odot}~{\rm yr^{-1}}$.
Since the sub-mm bump is almost completely from the innermost region ($r< 30~\rg$, as will be shown later),
Fit 3 is made by taking into account only the sub-mm bump in order to acquire some rough idea about the innermost accretion flow.
The data of sub-mm bump ranges from $10^2$~GHz to IR band.
In this range, the SED is dominated by the innermost region and meanwhile as much as possible data points are included.

\section{Results}

The contours of MCMC sampling, or the posterior probability density distributions, are shown in Figures~\ref{fig:triangle1}-\ref{fig:triangle3}.
From Figure~\ref{fig:triangle1} it can be seen that the parameters are strongly correlated.
It is a natural result because the three parameters are all related with density, 
from which the observed flux arises. 
In other words, the three parameters determines different profiles of density distribution.
When $s$ increases, the outflow is strong and $\dot{m}_0$ has to be higher to keep the luminosity.
This in turn leads to an increase in density at large radius, further requiring a smaller fraction of non-thermal electrons.

For Fit 1, the result of 1-$\sigma$ level is 0.37(0.33,0.39), 1.5 (1.2, 1.8) $\times 10^{-7}M_{\odot}~{\rm yr^{-1}}$, 2.2(1.9, 2.7)\%,
for $s$, $\dotm_0$, and $\eta$, respectively.
In Figures~\ref{fig:triangle2} \& \ref{fig:triangle3}, the histograms are not symmetric, with the most possible value of certain parameter locating at the lower limits.
Due to the non-symmetricity of the contours of probability density, we do not concern the mean value for Fits 2 \& 3, instead we select the values corresponding to the highest peaks in the histograms as the best fit values.
The results are listed in Table ~\ref{tab:para}.
It is noteworthy that the value of $s$ are very different for the different fits, which are about 0.37, 0.6, and 0.0, respectively.

The contributions of the thermal and non-thermal electrons inside and outside $30~\rg$, as well as the total SED, are shown in Figure~\ref{fig:sedfit}. As an example, only Fit 1 is shown.
The following points can be seen.
For the most luminous part of the SED, the sub-mm bump is dominated by the synchrotron emission of the thermal electrons within $30~\rg$.
In the low-frequency ($\nu<10^{10}~{\rm Hz}$) band, the contribution of non-thermal electrons from larger radii becomes significant.
Around $10^{10}-10^{11}~{\rm Hz}$ or cm band, the emission from non-thermal electrons is almost comparable to that from inner thermal electrons.
The X-ray emission at 5~keV is dominated by the synchrotron emission of the inner non-thermal electrons,
and the hard X-rays around $\sim 100~{\rm keV}$ mainly emit from bremsstrahlung of the outer thermal electrons.

The SED of all the three fittings are shown in Figure~\ref{fig:compa}, with the left panel showing the multi-band SED and the right panel zooming in the sub-mm bump.
The differences between the fittings are easy to understand.
The thermal synchrotron bump are almost the same for the three fittings because the densities and temperature are almost the same in the innermost region.
For Fit 2, the stronger outflow or flatter density distribution leads to higher density at larger radius, as a result a smaller $\eta$ is needed to fit the cm and low-frequency radio data.
The smaller $\eta$ then leads to less non-thermal electrons in the innermost region and consequently weaker 5~keV X-ray emission.
Moreover, the higher density at large radius can also lead to stronger emission in the hard X-ray band.
For Fit 3, because the density decreases with radius quickly, even $\eta$ is large, the density of non-thermal electrons is still too small to explain the cm and low-frequency radiation.
Let alone the value of $\eta$ in Fit 3 can only be slightly larger than that in Fit 1 if the limitation of 5~keV X-ray data is included.

Comparing all the fittings with the observational data, the following points can be seen. 
Fit 2 can explain the SED well, but it slightly underestimates the peak of the sub-mm bump.
Fit 3 can explain the bump quite well, but underpredicts low-frequency radio data($\nu < 10^2$~GHz) and slightly overpredicts the X-ray data.
Fit 1, intermediate between Fit 2 and Fit 3, explains the SED best generally.

Although both Fits 1 and 2 can explain the SED well, 
it is indispensible to check whether they are consistent with \citet{Roberts2017}, which is fundamental to this paper.
The photon index at 5~keV is given in \citet{Roberts2017}, i.e., $\alpha \sim$4.8(3.5, 7.5).
Both fittings are consistent due to the large uncertainty.
On the other hand, we check the density and accretion rate at $10^3~\rg$.
The density distributions of the fittings have been shown in Figure~\ref{f:density}, and the value of $\dotm_0$ is listed in Table~\ref{tab:para}.
The densities of different models are similar in the innermost region due to the sub-mm bump, 
but they are very different at larger radii.
For Fit 1, $\dotm_0$ is a few times lower and the density $n$ is a bit higher, 
but roughly speaking the result is consistent.
For Fit 2, the density at the outer boundary is almost one order of magnitude higher, which is out of the uncertainty.
So Fit 1 is preferred, though the value of $s$ in Fit 2 agree very well with the numerical simulations.

The fitting results, especially the value of $s$, are not significantly affected by the value of the fixed parameters, i.e., $\beta$, $\delta$, and $p$.
For $\beta$ and $\delta$, to some extent, they degenerate with $\dotm_0$, and so are already incorporated.
For $p$, because of the constraints from the flux ratios between radio bands and X-ray band, its value is unlikely to change much (Y03).

To further check whether our results are reasonable or not, in the following four paragraphs, we check the value of the parameters by comparing with previous work and observations.
Compared to previous results, most of the parameters are of reasonable value. 
Although the value of $\eta \sim 2\%$ seems a bit high,
it is still an acceptable value, particularly in the inner region.
This is because the fraction of non-thermal electrons could be as high as 10 percent, as shown by the studies about the particle acceleration mechanisms \citep[e.g.][]{Guo14,Guo15,Ball2016}.

Compared with Y03, the sub-mm bump is lower and flatter.
This is because their modeling is dominated by the data points at 230~GHz and 340~GHz, which are much higher than the interpolation of the other radio data, 
and are likely contaminated by diffuse emission from the surrounding region when the resolutions of early observations are poor.
The new observational data by ALMA also considerably lower than those two data points.
Due to their adopted higher fluxes, the density inferred in Y03 is also higher than that in our fittings, as shown in Figure~\ref{f:density}.

From the global solution of the basic equations of the accretion flow, the distribution of density and temperature can be calculated, and from the parameter $\beta$, which is assumed to be about 10 in our calculations, we can get the strength of the turbulent magnetic field.  
Integrating the product of the density and the magnetic field over the radius, the upper limit of the RM along the equatorial plane can be obtained. 
The upper limit of RM in Fit 1 is $\sim 2 \times 10^6$~${\rm rad~m^{-2}}$. 
This is not much greater than what have been observed, which is about $5\times 10^5$~${\rm rad~m^{-2}}$ \citep{Marrone07}.
Taking into account the reversals of the magnetic field along the line of sight \citep[Y03,][]{Li16}, our results could be easily reconciled with the observations.
As also pointed out by \citet{Marrone07}, the estimated upper limit of accretion at $\sim 100~\rg$ from observations is about $2 \times 10^{-7}M_{\odot}~{\rm yr}^{-1}$.
Our results agree with this limit quite well.

\begin{figure}
\centerline{
\epsfig{figure=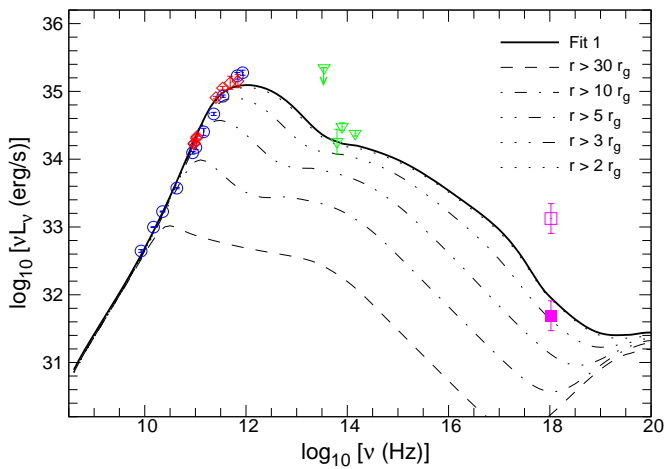,width=0.5\textwidth,angle=0}}
\caption{ 
The contribution of the innermost accretion flow to the SED in Fit 1.
}
\label{fig:componentr}
\end{figure}

Great efforts have been paid to the innermost region of Sgr~A*, trying to reveal the shadow of the horizon \citep[e.g.,][]{Bower04,Shen05,Doeleman08,Falcke18}.
Although the radio emission at cm band suffers from the broadening due to the scattering by ionized gas along the sight line \citep{Shen05,An05,Bower06,Sicheneder17}, the observed size of Sgr A* at 1.3 mm is most likely intrinsic, which corresponds to a physical radius to be about $7.5~\rg$ \citep{Doeleman08}.
The observed emission at different frequencies are dominated by different annuli, as shown in Figure~\ref{fig:componentr}.
At 1.3 mm, the emission peaks at by the annulus arround $r=6.8~\rg$, which agrees with the observed result within the uncertainty.

\section{Discussion}

\subsection{Limitation to the outflow}

In Fit 1, the value of $s$ is a bit larger than that of Y03, but is still smaller than the numerical simulation result. 
How to reconcile the numerical simulations with the observational result is therefore a concern.
As mentioned in \citet{Yuan2014}, this difference may be related to some detailed physical processes that have not been included, such as the low angular momentum of the accretion flow \citep{Bu13} or the dynamical importance of thermal conduction \citep{Johnson07}.
But no detailed explanation is given why these processes could reconcile. 
Here we propose that the smaller value of $s$ may in fact support the numerical simulations, 
indicating the domination of the inflows over the outflows in the region very close to the central black hole.

In Fit 1 all the multi-band SED data are included during fitting, while in Fit 3, only the sub-mm bump data are taken into account.
Considering the most significant part of the SED, i.e., the sub-mm bump, is fitted in both fittings,
if the value of $s$ were about the same at different radii in the true accretion flow, the fitting result of Fit 3 should agree with Fit 1 within uncertainty.
However, the $s$ values from the fits are too different to be reconciled with uncertainty.
The difference can probably only be explained by variation of $s$, with $s\sim0$ in the innermost region and $s$ being larger at larger radius.
In fact, such variation of $s$ is just what have been found in numerical simulations \citep[e.g.,][]{Stone99,Yuan12b,Narayan12,McKinney12,Li2013}.
According to the numerical simulation, there is a turning point/radius, 
inside which the outflow is dominated by the inflow with $s\sim 0$ while outside outflow is significant with $s\sim 0.5-1$ \citep[e.g. Fig.5 in][]{Yuan2015}.
At present, all the fittings and modelings consider $s$ to be fixed in radius, and can in some sense be thought of as the average of $s$ between the regions inside and outside the turning radius.
Therefore the value of $s$ from SED fitting/modeling is smaller than the value of numerical simulations, $0.5-1$, because of the effect of average. 

Compared to Fit 1, the limitation of the accretion rate at the outer boundary in Fit 2 enhances the weight of the outer region. Therefore the value of $s$ is larger. 
However, the value of $s$ in Fit 2 should still be smaller than the true value in the outer region due to the same average effect. 
So we expect the true outflow index of the accretion flow in Sgr A* should be in the range of $\sim 0.6-1$, which could be consistent with the results of numerical simulations \citep[e.g.][]{Yuan12, Yuan2015}.

Since the outflow is weak in the innermost region, we hereafter refer the accretion flow with a turning point as weak-outflow model.
To be consistent with \citet{Wang2013,Roberts2017}, the outflow index is supposed to be $s\sim 1$ in the outermost region.
Combining the advantages of Fits 2 and 3, such a weak-outflow model for the inner region could explain the SED even better than Fit 1.
Like Fit 3, the weak outflow in the innermost region can fit the sub-mm bump better,
and like Fit 2, higher density in the outer region requires a less fraction of non-thermal electrons, i.e., smaller $\eta$, which could explain the X-ray data better and seems more reasonable.

For the weak-outflow model, given the values of $s$ in the inner and outer region, it is possible to obtain the position of the turning point from SED fitting.
However, with the change of $s$, the density could not be described by a simple power-law.
Thus, we intend to develop such a model in a future work.
Here as a check, we extend the result of \citet{Roberts2017} inward, i.e. $\dotm(r)=10^{-6}(r/10^3~\rg)M_{\odot}~{\rm yr}^{-1}$,
and calculate the radius at which it intersects with Fit 1.
The intersection radius should be just a bit larger than the turning point,
since Fit 1 overestimates the accretion rate in the innermost region due to the effect of averaging. 
Substitute the best fitting parameters of Fit 1, the intersection radius of the accretion rate is found to be $r=49~\rg$.
Considering the uncertainties of the fitted parameters, the turning radius ranges from $31~\rg$ to $77~\rg$. 
The range agrees quite well with the turning radius obtained from the numerical simulations, being in the range between $\sim 30~\rg$ and $\sim 90~\rg$ \citep[e.g.,][]{Stone99,Yuan12b,Narayan12,McKinney12,Li2013}.

In addition to the results above that suggest $s$ should vary with radius,
there remain other parameters that are possible further improve the fitting.
In our above calculations, the fraction of viscous dissipation heating electrons, $\delta$, is assumed to be a fixed value.
However, its value is still unclear. Recent work by \citet{Chael18} shows that if the electron heating is due to magnetic reconnection, $\delta$ can then be the same everywhere, but if the heating is due to the dissipation of turbulence, $\delta$ would then increase with decreasing radius. 
To consider the potential effect of turbulence heating, we recalculate the spectra with $\delta=\delta_0+0.5/r$, where $\delta_0=0.1$ is the assumed value at large radius, and the value of $\delta$ approach to 0.6 close to the black hole.
And then we fitted the SED with a new tabulate of the model spectra.
Indeed, the fit at the synchrotron peak is improved a little.
Another potentially important parameter is the ratio of gas pressure to magnetic pressure, $\beta$.
Recent GRMHD numerical simulations by \citet{Ball2016} indicate that this value of $\beta$ may decrease at low radii ($r<20~\rg$).
If this is the case, the synchrotron emissivity should then increase in the innermost region, 
which may improve the fit to the sub-mm peak.

A prominent issue that needs to be addressed is the estimation of the quiescent emission itself, as noted in Section~\ref{sec:methods}.
Separating the X-ray flares from the quiescent emission is perhaps the easiest, 
as they are the most distinct.
Their flux can be up to 100 times greater than the quiescent rate.  
The X-ray flares are also narrow temporally, with a typical duration of $10^3$~seconds.  
However, at longer wavelengths, the flare profile tend to become much broader, 
and the emission typically fluctuates by only a factor of a few at the mm band, for example. 
Therefore, it becomes difficult to define the quiescent state due to their temporal broadness and flare weakness \citep{Genzel10}.  
Since the mm/sub-mm data points in the SED were estimated through the mean flux, it is possible that the true quiescent flux is several tenths percent lower.  
In our model, this lower flux would naturally be compensated by a decrease in density at the inner radii through a stronger outflow solution.
So more observations and studies on the quiescent radio emission in general are still needed.  
It should be noted that the upcoming sub-mm VLBI data of Sgr A* obtained by the Event Horizon Telescope (EHT) will resolve the innermost region of the accretion flow to a resolution of $\sim 25~{\rm \mu as}$, i.e. $\sim 5~\rm{r_g}$,
which can not only constrain the quiescent radio emission better, but also directly test the weak outflow in the innermost region.

Another issue arises from the simplification in our calculation related with the relativistic potential. 
The dynamics of the accretion flow is calculated based on the pseudo-potential instead of the Schwarzschild metric. And the radial distribution of the density and the electron temperature are then modified before calculating the spectra.
This is not precise, and future self-consistent calculations could give more conclusive result.

As a first step of the multi-band SED fitting, the present paper does not include some subtle,
but potentially important effects of BH spin and general relativity.
First, our model assumes that the BH is non-rotating \citep[see also][]{Yuan06}.
For fast spin BH, the gravitational potential is steeper and consequently the flux could be higher. 
The uncertainty is estimated to be less than 30 percent according to the results of \citet{Manmoto00}, which calculated the multi-band SED in the Kerr metric. 
In fact, this uncertainty does not affect our result that much.
On the one hand, because of the degeneration among the parameters, this uncertainty should have been included in $\dot{m}_0$ \citep{Feng16}. 
On the other hand, the spin of the BH in the Galactic Center should be slow, otherwise the jet power would be expected to be strong enough \citep{Blandford77}, which seems to be inconsistent with the existing observations.
Second, the gravitational redshift, as well as the Doppler effect, should be considered when the emission is close to the BH \citep[e.g.,][]{Cunningham75,Fabian00,Reynolds03}. 
From Figure~\ref{fig:componentr}, it can be seen that the broadening effects should be limited in the range from $\sim 10^2$ to $10^3$ GHz ($r < 5~\rg$), where the observational data suffer relatively high uncertainties.
So it is still difficult to constrain the theoretical models with the GR effects at present.
But, in the future, with the increasing precision of the sub-mm radio observations, the models that include the above GR effects may become necessary.

\subsection{On the origin of non-thermal electrons}
The origin of non-thermal electrons in the quiescent Sgr A* remains to be studied.
\cite{Kowal2012} explored how different mechanisms of particle acceleration affect the energy of particles.  Specifically, they show how the energy of a particle increases as a function of time with respect to the acceleration process.  For the fiducial flare particle acceleration mechanism, magnetic reconnection, particles are efficiently accelerated (E $\propto$ t$^{1.43}$).  
Contrastingly, when reconnection becomes unimportant and turbulence drives particle acceleration through a second order Fermi process where E $\propto$ t$^{0.66}$.  To first order, this decreased efficiency appears to naturally result in a spectrum of powerlaw index of the accelerated electrons $\sim3.5$ when considering the powerlaw index of flare emission ($\sim2.6$), which would be in broad agreement with the results of Y03 and \citet{Roberts2017} that we followed in this paper.

Moreover, according to our calculations, as shown in Figure~\ref{fig:sedfit}, the radiation at the sub-mm radio and X-ray bands are dominated by thermal and non-thermal electrons in the innermost region, respectively.
Since the sub-mm and X-ray radiation are from the same region, given the energy spectrum of non-thermal electrons, the flux ratio between sub-mm and X-ray bands could limit the fraction of thermal electrons to non-thermal one.
According to our previous results, i.e., the value of $\eta$ of the weak-outflow model should be smaller than that of Fit 1, we expect $\eta < 2.2$\%.
Similarly, the low-frequency radio emission and hard X-rays are all from the outer region, the correlation between them should also be useful to understand the acceleration.

In our model, the fraction of non-thermal electrons is supposed to be fixed at different radii.
In reality, the value of $\eta$ could be different due to the different physical condition.
As shown in Figure~\ref{fig:sedfit}, the emission around a few~GHz and 5~keV is dominated by non-thermal electrons at $\sim 100 \rg$ and $\sim 10 \rg$, respectively.
Consequently, the study of the long-term correlation between a few~GHz and 5~keV bands may help us to understand the variation of $\eta$ at different radii.

\section{Summary}

The multi-band SED of the inner accretion flow ($r<10^3~\rg$) around Sgr A* is investigated for the first time. This analysis becomes possible because of the new X-ray analysis by \cite{Roberts2017}.  
Three different fittings are made to explore the inner region of the hot accretion flow.
The low-frequency radio data, sub-mm bump/NIR data, and the X-ray data of the SED 
can be reasonably well explained by the non-thermal electrons in the outer region ($\sim 30~\rg< r < 10^3~\rg$), 
thermal electrons in the innermost region ($r<\sim30~\rg$), the non-thermal electrons in the innermost region, respectively.
In the cm band, $10^{10}-10^{11}$~Hz, both the thermal and non-thermal electrons are important.
Considering the limitations of the accretion rate and emission region,
the fitting results of $s$ are very different,
which indicates that in the innermost region the outflow should be dominated by the inflow.
The weak-outflow model can naturally explain why in previous SED modelings the value of $s$ was smaller than that obtained from numerical simulations.
This model is also able to explain the SED better than the model with fixed $s$.
Our results support the numerical simulations of the hot accretion flow in the inner most region \citep[e.g.,][]{Stone99, Yuan12b,Narayan12, McKinney12,Yuan2015}.

Moreover, considering the contribution of non-thermal electrons to the multi-band SED, the studies about the correlations between sub-mm bump and X-ray emission may help us to understand the particle acceleration in the innermost region of the accretion flow.

\section{Acknowledgements}

We would like to thank Feng Yuan for his thoughtful comments.
We would also like to thank Fu-Guo Xie, De-Fu Bu and Qing-Wen Wu for their helps.
MRY is supported by the National Natural Science Foundation of
China under grants U1531130, 11333004, the Fundamental Research Funds for the Central University under grants 20720150024, and the
Natural Science Foundation of Fujian Province of China under grant 2018J01007.
YPL was also sponsored by the National Natural Science Foundation of
China under grants 11703064 and the Shanghai Sailing Program (No. 17YF1422600).

\bibliographystyle{aa}
\bibliography{sgras_fitting}

\end{document}